# Sound Localization and Separation in Three-dimensional Space Using a Single Microphone with a Metamaterial Enclosure

Xuecong Sun[1,2], Han Jia[1,2,3*], Zhe Zhang[1,2], Yuzhen Yang[1,2], Zhaoyong Sun[1], and Jun Yang[1,2,3*]

Dr. X. C. Sun, Prof. H. Jia, Dr. Z. Zhang, Dr. Y. Z. Yang, Dr. Z. Y. Sun, Prof. J. Yang
Key Laboratory of Noise and Vibration Research,
Institute of Acoustics, Chinese Academy of Sciences,
Beijing 100190, People's Republic of China
E-mail: hjia@mail.ioa.ac.cn; jyang@mail.ioa.ac.cn

Dr. X. C. Sun, Prof. H. Jia, Dr. Z. Zhang, Dr. Y. Z. Yang, Prof. J. Yang
University of Chinese Academy of Sciences
Beijing 100049, People's Republic of China

Prof. H. Jia, Prof. J. Yang
State Key Laboratory of Acoustics
Institute of Acoustics, Chinese Academy of Sciences
Beijing 100190, People's Republic of China

## Abstract

Conventional approaches to sound localization and separation are based on microphone arrays in artificial systems. Inspired by the selective perception of the human auditory system, we designed a multi-source listening system which can separate simultaneous overlapping sounds and localize the sound sources in three-dimensional space, using only a single microphone with a metamaterial enclosure. The enclosure modifies the frequency response of the microphone in a direction-dependent manner by giving each direction a characteristic signature. Thus, the information about the location and the audio content of sound sources can be experimentally reconstructed from the modulated mixed signals using a compressive sensing algorithm. Due to the low computational complexity of the proposed reconstruction algorithm, the designed system can also be applied in source identification and tracking. The effectiveness of the system in multiple real-life scenarios was evaluated through multiple random listening tests. The proposed metamaterial-based single-sensor listening system opens a new way of sound localization and separation, which can be applied to intelligent scene monitoring and robot audition.

## 1. Introduction

Sound localization and separation are both fundamental tasks in signal processing with a wide range of applications, including robot audition, speaker recognition and medical detection.[1–3] Given their importance, these two tasks have been extensively studied in the signal processing field for decades. In artificial systems, sound localization and separation are typically performed using two or more microphones, known as the microphone array. However, the accuracy of localization and separation of the microphone array is fundamentally limited by the number of the microphones and the physical size of the array.[4–7] Considering that such microphone arrays could be cumbersome to mount and maneuver, their utility is much reduced in certain situations.[8]

A lot of effort has been made in overcoming the shortcomings of the microphone array by developing biomimetic systems.[9,10] In biological organisms, such as humans, sound localization and separation can be performed with high accuracy using just two ears, and sometimes even just a single ear suffices. Accomplishing such tasks is primarily due to several key features of the human auditory system. First, unlike common microphones, the eardrum is not exposed directly to the air. Therefore, the sound is scattered by the head, torso and the ear pinna before its arrival at the eardrum. Due to the irregular spatial structure of the scatters, the sound measured in the eardrum changes as a function of the source's direction, referred to as the head-related transfer function, which is the basis of monaural localization and separation.[11–13] Second, accomplishing this task requires the prior knowledge of received sounds.[14] In fact, the ability of humans to estimate the direction of a sound monaurally is contingent on their familiarity with it. We always estimate the direction based on our years of experience about what sounds are likely.[15–17]

Researchers have attempted to replicate these bionic features in artificial listening systems in an effort to achieve monaural sound localization and separation.[14,18–21] To design such a system, one has to deal with two important issues. First, a learning mechanism should be introduced into the system, which is embodied in the reconstruction algorithm, such as compressive sensing and Hidden Markov Model.[22–24] Second, an anisotropic scattering mechanism should be designed in order to provide monaural cues, such as placing irregular scatters surrounding the microphone.[19,21] Among the numerous approaches in recent years, acoustic metamaterials have served as an optimal choice for use in designing an anisotropic scatter mechanism. Acoustic metamaterials have been proven to be effective in modulating both the phase distribution and the amplitude distribution of acoustic signals.[25–30] Combining acoustic metamaterials and compressive sensing, a

Helmholtz-based device was used to localize known noise sources, and a space-coiling anisotropic metamaterial was designed to demonstrate a single-detector acoustic camera.[18][20] However, both of these studies primarily focused on the two-dimensional(2D) space. Additionally, an extra microphone is needed to obtain the phase information of the measuring signals and the reconstruction algorithm relies on the full complex-valued spectrum, which limits their real-time performance and application.

Taking into consideration the current problems and the previously reported designs, in this study we developed a metamaterial-based single-microphone listening system (MSLS), which can localize and separate multiple sound signals from an overlapping signal in three-dimensional(3D) space. A 3D metamaterial enclosure (ME) is designed to provide monaural cues to the inversion task by coding sound signals as a function of the source direction. The information regarding the sound sources could be reconstructed from the modulated mixed signals utilizing the compressive sensing framework. During signal processing, a joint algorithm of variable sparsity principal component analysis and orthogonal matching pursuit (VSPCA-OMP) is used to solve the multi- source listening problem, which just relies on the power or magnitude spectrum of the signal. Due to the low computational complexity and high robustness, the system not only realizes the sound localization and separation, but also has a great real-time performance in source identification and tracking. We experimentally demonstrate the performance superiority of the MSLS using challenging sources in real-life scenarios. These results promise a wide range of potential applications for our proposed system, such as intelligent scene monitoring and robot audition.

## 2. Theory and Design

### 2.1. Metamaterial Enclosure Design.

Inspired by the frequency-dependent filtering mechanism, we designed the microphone enclosure with carefully engineered metamaterials in order to achieve dispersive frequency modulation. The proposed ME is shown in **Figure 1**(a), which is composed of three-layer hemispherical shells. Multiple holes are randomly drilled on the hemispherical shells. Eight transverse plates and 16 longitudinal plates are also randomly inserted between these shells to divide the hemisphere layers into 24 different cavities. A single microphone is placed at the center of the hemisphere to receive signals from different directions modulated by the ME.

The ME can be regarded as the combination of multiple acoustic channel modules (ACMs) towards to different directions. One of the ACMs is marked with a red frame in Figure 1(a). The ACM can be regarded as a second-order acoustic filter, which includes three layers of perforated plates and two cavities. The geometric parameters of the ACM, including the volume of the cavities, the filling ratio, and the location distribution of the holes, would directly affect the frequency response (see S1 of the Supporting Information). During the design process, holes with different radii are randomly distributed on the hemispherical shells and the plates are also randomly inserted between the hemispherical shells. Therefore, the volumes of the cavities, the filling ratio, and the location distribution of the holes are all random for these ACMs.

The structural randomness of the ACMs essentially leads to the stochastic frequency response. Considering that the ME is composed of different ACMs towards different directions, the frequency response of the ME would be directionally dependent. As a result, the original omnidirectional measurement mode of the single microphone is modified by the randomized modulation from the ME.

The frequency responses below 5 kHz of the designed ME in four different directions are calculated using the finite element method (FEM), which are shown in Figure 1(b). Although the frequency responses in the low frequency band look similar for the finite geometric size of the ME, one observes that the frequency responses drastically vary with direction in a wide frequency range as a result of the modulation by the ME. Here, we choose the coherence $\mu_{ij}$ to quantitatively evaluate the independence of the frequency responses in different directions. The coherence between two responses $h_i$ and $h_j$ is defined as $\mu_{ij} = \langle h_i, h_j \rangle / |h_i||h_j|$, which represents the overlap between the two frequency responses. Due to the omnidirectional measurement mode of the microphone, the coherences between two different directions would be close to 1 without the ME. After the modulation of the ME, the frequency responses vary from direction to direction and the coherences would also be reduced. The coherences between the four frequency responses mentioned above are shown in Figure 1(c). It can be seen that the coherences between two different directions are all less than 0.91. The above results demonstrate that the variation of the geometric parameters of the ME is large enough to ensure that the frequency response is sensitive to direction.

## 2.2. The framework of Signal Processing Algorithm.

Compressive sensing framework is employed to afford the MSLS with learning capabilities. In the framework of compressive sensing, the proposed listening system can be described with a general model:

$$y = \mathbf{A}s \tag{1}$$

where $y$ is the vector form of the measured data (observation vector); $s$ is the sparse representation of the signals (object vector), which can be estimated to reconstruct the original signals; Matrix $\mathbf{A} = [a_1, a_2, ..., a_Q]$ is a $P \times Q$ matrix (measurement matrix) determined by the ME and the signals, which is constructed to store the prior knowledge of possible sounds.

The concept schematic of the data collection and processing for the MSLS is shown in **Figure 2**. The sound waves are first emitted from the sources and then they propagate to the surface of the ME. Then they are modulated by the ME and collected by the microphone at the center of the enclosure as a single mixed waveform. In the free field, the mathematical model of the interaction between the sound waves and the ME can be expressed as

$$f_m(\omega) = f_s(\omega) \circ h_i(\omega) \tag{2}$$

where $f_m(\omega)$ is the spectral amplitude of the signal collected by the microphone; $f_s(\omega)$ is the spectral amplitude of the original signal; $h_i(\omega)$ is the frequency response of the MSLS, which is a direction-related function; "∘" is the Hadamard product, which indicates element-wise multiplication[31]. Therefore, the measured signals will be different even for the same original signals coming from two different directions.

In order to reconstruct the location and the audio content of each source from the mixed waveform, the signal processing has been divided into two steps: training and testing. During the training procedure, the training signals are collected one by one to be used as the training set. After a series of signal preprocesses (see the Experimental Section), the measurement matrix **A** can be experimentally constructed (see the S2 of the Supporting Information).[32] Here, the matrix **A** is constructed as a real matrix, i.e., we only need the spectrum amplitude of the measuring signals, without phase information. Therefore, the signals are collected only by the MSLS without the additional referenced microphone. As a result of this, the computational complexity of signal processing has also been greatly reduced. After training, a sequence of sounds is randomly chosen from the audio library and simultaneously generated from the sources in order to evaluate the performance of the MSLS. This process is described as testing, where the observation vector *y* is obtained. The observation vector *y* can be regarded as the weight sum of certain columns of matrix **A**, which correspond to the nonzero elements of the object vector *s*. Therefore, the object vector *s* can be recovered using algorithmic approaches based on the matrix **A** and vector *y*. Then both the locations and the audio contents of the active sources can be reconstructed through estimating the object vector *s*.

The two major algorithmic approaches to sparse recovery are the L1-minimization method and the iterative method. Here, Orthogonal Matching Pursuit (OMP), a kind of iterative method, is applied to the sparse recovery.[33] The OMP performs faster and is easier to implement than the L1-minimization method.[34] It only requires $k$ iterations to recover a $k$-sparse signal. For a measurement matrix $\mathbf{A} \in \mathbb{R}^{M \times N}$, each iteration includes two steps: First, a multiplication between a $N \times M$ matrix and a $M \times 1$ vector requires to be calculated; Then, a least squares problem in $M \times k$ dimensions needs to be solved based on the calculation results in the previous step. However, the condition of OMP on measurement matrix $\mathbf{A}$ is more restrictive than the Restricted Isometry Condition. Therefore, we need to further reduce the correlation of the columns of the matrix $\mathbf{A}$ by transforming both the training samples and the testing samples. Considering that one testing sample can be regarded as a linear superposition of several training samples, the transformation should be a linear transformation. Here, we propose an improved algorithm called variable sparsity principal component analysis (VSPCA).[35] Compared with the principal component analysis (PCA), the sparsity $k$ is introduced to the normalization process to ensure that all of the samples can be mapped to the same space. This allows us to reduce the dimension for each signal sample while still capturing most of their variability according to the sparsity (details concerning the VSPCA algorithm can be found in the S3 of the Supporting Information). In order to demonstrate the effectiveness, the proposed VSPCA was used to transform the frequency responses in Figure 1(b). The coherences after transformation are calculated and shown in Figure 1(d). It can be seen that the coherences are greatly reduced compared with Figure 1(c). Therefore, the columns of the matrix $\mathbf{A}$ become more independent, which further improves the reconstruction accuracy.

## 3. Experiment

**3.1. Sound Localization and Separation.**

The effectiveness of the MSLS was evaluated using listening tests, which should be able to identify the locations of the activated sources while simultaneously segregating the overlapping signals. The listening tests were conducted in a semi-anechoic room and the experimental setup is shown in **Figure 3**(a). The ME, whose enlarged photo is shown in Figure 3(b), was deployed on the center of the floor. A microphone was placed in the inner center of the hemisphere shell. The bottom of the inner hemisphere was sealed using the absorption cotton to reduce the reverberation. Sixteen speakers were used as sound sources, of which the locations are schematically shown in Figure 3(c). These speakers formed 2 rings, with radii of 2.5m and 4.5m, which evenly distributed around the MSLS. To evaluate the effectiveness of the MSLS in 3D space, the speakers in the inner ring were deployed on the floor, while the speakers in the outer ring were mounted on the stands with a height ranging from 90cm to 100cm.

The audio library used in the tests contained 6 typical sound signals often heard in the street, including sounds of police car, backing car, ambulance, car whistle, fire engine, and bicycle bell. All of these signals are wideband and most energy is distributed below 5 kHz (the spectrograms and power spectral densities of these signals are provided in **Figure S5** of the Supporting Information). The training process should be experimentally performed in advance. Each signal in the audio library was successively emitted from 16 different locations and collected by the microphone at the center of the enclosure to obtain the measurement matrix **A**. During the testing process, one or more of the speakers from the 16 different locations were randomly selected as activated sources. The audio contents were also randomly selected from the audio library and simultaneously emitted by the activated sources. The test signals from multiple sound sources were mixed in the transmission process and then collected by the microphone. After a series of signal processing, the object vector $s$ could be reconstructed. The reconstruction results are evaluated by the success rate, which is defined as $\alpha = n/k$. Here, $n$ is the number of the recognized sources, which means both the location and audio content of the sources were reconstructed successfully; $k$ is the total number of the activated sources, also known as sparsity.

The sparsity $k$ in the tests varied from 1 to 5. We conducted 100 random experiments for each sparsity. The experimental results are arranged by the number of activated sources as well as by the audio contents, which are shown in **Figure 4**(a) and (b) respectively. In Figure 4(a), the distribution of the reconstruction results for each experiment is displayed by different colors. As $k$ increases, $\alpha$ gradually decreases. This is because the projections of high sparsity signals on the dictionary may not be unique. Despite this, the results of 100 random experiments indicate that the average recognition ratio $\alpha$ is greater than 90% when the number of activated sources $k$ is no more than three. At the same time, the average recognition ratio of the MSLS is still close to 70%, even if the $k$ is up to five.

Figure 4(b) gives the detailed results of the 6 kinds of audio contents. The average success rates almost exceed 90% for all of the audios in cases with a small sparsity *k*. As *k* increases, the accuracy of the reconstructions varies for different audios due to different frequency components. Due to the size limit, the metamaterial-based enclosure is more capable in modulating the high frequency waves. The audio samples with more energy in the higher frequency range are more likely to be reconstructed. The experimental results also show that the error is somewhat higher for the audios whose energy concentrates on some narrow frequency ranges, such as the siren sound of the police car. This is because the spectral shaping of these audios is not large enough. Indeed, psychological studies have shown that humans also cannot localize narrowband signals.[36]

The proposed MSLS is applicable to many other scenarios. We also examined the performance of the MSLS in other 4 scenarios, including home, animal farm, speech and concert. For each scenario, we still selected 6 typical sounds to conduct the listening tests (the spectrograms and power spectral densities of the signals are provided in **Figures S6~9** of the Supporting Information). The results are similar to the results we obtained from the tests in the street scenario, which are shown in **Table 1** (details concerning the results are provided in **Figures S10~13** of the Supporting Information). When the number of the activated sources is no more than 3, the average success rate is greater than 90%. Moreover, in order to validate the effectiveness of the MSLS in a larger corpus, another listening test containing 30 command words from the Speech Commands dataset was performed.[37] The results are shown in the last column of **Table 1**, which have the same variation trend of reconstruction accuracy. The results well demonstrate the stability of the MSLS in the case of larger audio numbers (details concerning the results are provided in **Figure S14** of the Supporting Information). In order to further examine the 3D spatial resolution, we also conducted the listening tests in pitching direction, which can be found in Section S4 of the Supporting Information. In addition, we made videos in which a portion of the tests were recorded in order to transform the listening tests into dynamic visualizations (see Supporting Videos 1 and 2). The above results demonstrate that the MSLS is capable of identifying locations while simultaneously separating audio contents from the mixed signals in 3D space, highlighting the wide range of potential applications of our system.

**3.2. Source Tracking and Identification.**

Owing to the real-time applicability of the VSPCA-OMP algorithm, the MSLS can also be applied in source tracking and identification. In order to evaluate this, we again chose the street scenario. To simulate several cars moving in the street, we kept the activated speakers moving artificially and collected the signals during the tests. The trajectories of the speakers are shown in **Figure 5**. The first listening test simulated an ambulance moving around the system, which is schematically illustrated in Figure 5(a). In the second listening test, two activated speakers which played sounds of backing car and fire engine, moved simultaneously as shown in Figure 5(b). Owing to the low computational complexity of the reconstruction algorithm, the tracking for multiple objects had been realized within one second. At the same time, the objects were also accurately identified. The results of the two tests can be found in the Supporting Videos

3 and 4. In addition, all of the sound signals, including unavoidable noises produced by moving the speakers, were monitored and processed during the tests. And there was no denoising algorithm during the signal processing, which shows a certain anti-noise performance of the MSLS. The above reconstruction tests demonstrate that the MSLS can identify multiple sources and track them quickly and accurately.

## 4. Conclusion

In this paper, we present a 3D listening system using a single microphone in combination with the ME, which functionally mimics the listening capability of the human auditory system. A well-designed ME is used to break the original omnidirectional measurement mode of the single microphone. And a joint algorithm VSPCA-OMP is presented to solve the multisource listening problem, which has the advantages of low computational complexity and good real-time performance. Listening tests were conducted in several common scenarios, in which 16 speakers emitted sound signals from various directions in 3D space. The test results show that the MSLS is capable of localizing multiple sound sources while simultaneously separating audio contents from the mixed signals within one second. Due to the low computational complexity of the reconstruction algorithm, the proposed system can also be applied in source identification and tracking. We envision that the MSLS can be useful for multisource speech recognition and segregation, which is desired in intelligent scene monitoring and robot audition. In the future, denoising algorithm can be added in the signal processing procedure to enhance robustness in complex environments.

## 5. Experimental Section

*ME prototype:* The geometric structure of the ME is designed to be extremely irregular in order to make the received signals appear different depending on what direction it is coming from. Therefore, we introduce the randomizing algorithm to the design process. The designed ME is composed of three-layer hemispherical shells. The radii of the three hemispherical shells from the outer to the inner are 24 cm, 16.8 cm, and 7.2 cm, respectively; and the thickness of the shells is 1 cm, 0.7 cm, and 0.3 cm, respectively. Eight transverse plates and sixteen longitudinal plates, whose thicknesses are all 1 cm, are randomly inserted between these spherical shells to divide the hemisphere layers into 24 different cavities. Multiple holes are randomly drilled on the hemispherical shells, whose radii range from 0.3cm to 3cm. There is no overlap between any 2 holes, and the distance between them is larger than 0.5cm. The hole filling ratio of the perforated shells is set relatively high to ensure the high energy transmission and high quality of reception signals. Moreover, the variation of the geometric parameters is large enough to ensure different effective acoustic parameters in different directions.

The sample is fabricated using 3D printing technology with a precision of 0.06mm, which could ensure that there are no cracks in the connections between the shells and the plates. The material used for the sample is Lasty-KS, a type of UV curable resin, whose density is 1.13 g/cm$^3$. The acoustic impedance of the Lasty-KS is much larger than that of air, so the shells and the plates can be regarded as rigid for the sound wave.

*Experimental Setup:* The layout of the measurement system is shown in **Figure 6**. The signal receiver used in the listening tests is an omnidirectional microphone with diameter of 0.5 inch (B&K 4189-A-021). The signal acquired by the microphone is sent to an audio interface (Audient-iD14). Sixteen speakers (HiVi H5) are used as audio sources, which can be controlled independently through a DA converter (RME M-32 DA). Both the audio interface and the DA converter are connected to a computer. All of the emitted and received acoustic signals are controlled by this computer.

*Signal Processing:* The signal acquired by the microphone is a time-domain signal, while the reconstruction is based on the frequency domain information. The spectral amplitude of each frame of the signal can be calculated though framing and Short Time Fourier Transform (STFT). As shown in Figure 1(b), the similarity among these frequency responses in the low frequency band is higher than that in the high frequency band, which is not conducive to the signal reconstruction. Therefore, considering of the operational frequency range of the proposed system and the power spectral densities of the audio signals we used, the information of the signals between 100 Hz and 5000 Hz was preserved during the STFT. The signals were processed frame by frame both in the training and testing procedures. For the testing signals, each frame would correspond to a reconstruction result through the OMP algorithm. Then majority voting was used to get the final result based on these reconstruction results.

**Supporting Information**

Supporting Information is available from the Wiley Online Library or from the author.

**Acknowledgements**
This work is supported by the National Natural Science Foundation of China (Grant No. 11874383), the Youth Innovation Promotion Association CAS (Grant No. 2017029), and the IACAS Young Elite Researcher Project (Grant No. QNYC201719).

References

[1] Y. Tamai, Y. Sasaki, S. Kagami, H. Mizoguchi, in *2005 IEEERSJ Int. Conf. Intell. Robots Syst.*, IEEE, Edmonton, Alta., Canada, **2005**, pp. 4172–4177.

[2] F. Asano, M. Goto, K. Itou, H. Asoh, in *Seventh Eur. Conf. Speech Commun. Technol.*, **2001**.

[3] F. Ghaderi, S. Sanei, B. Makkiabadi, V. Abolghasemi, J. G. McWhirter, in *2009 16th Int. Conf. Digit. Signal Process.*, IEEE, **2009**, pp. 1–6.

[4] D. P. Jarrett, E. A. P. Habets, P. A. Naylor, *Theory and Applications of Spherical Microphone Array Processing*, Springer International Publishing, Cham, **2017**.

[5] B. D. Van Veen, *IEEE ASSP Mag.* **1988**, *5*, 4.

[6] J. Benesty, J. Chen, Y. Huang, *J. Acoust. Soc. Am.* **2008**, *125*, 4097.

[7] M. Brandstein, D. Ward, *Microphone Arrays: Signal Processing Techniques and Applications*, Springer Science & Business Media, **2013**.

[8] M. Park, S. Chitta, A. Teichman, M. Yim, *Int. J. Robot. Res.* **2008**, *27*, 403.

[9] H. Xu, X. Xu, H. Jia, L. Guan, M. Bao, *J. Acoust. Soc. Am.* **2015**, *138*, EL270.

[10] L. S. Smith, *Front. Neurosci.* **2015**, *9*, DOI 10.3389/fnins.2015.00398.

[11] S. R. Oldfield, S. P. A. Parker, *Perception* **1986**, *15*, 67.

[12] J. J. Rice, B. J. May, G. A. Spirou, E. D. Young, *Hear. Res.* **1992**, *58*, 132.

[13] M. Aytekin, E. Grassi, M. Sahota, C. F. Moss, *J. Acoust. Soc. Am.* **2004**, *116*, 3594.

[14] A. Saxena, A. Y. Ng, in *2009 IEEE Int. Conf. Robot. Autom.*, IEEE, Kobe, **2009**, pp. 1737–1742.

[15] B. A. Morrongiello, *J. Acoust. Soc. Am.* **1989**, *86*, 597.

[16] J. Sodnik, S. Tomazic, R. Grasset, A. Duenser, M. Billinghurst, in *Proc. 18th Aust. Conf. Comput.-Hum. Interact. Des. Act. Artefacts Environ.*, ACM, New York, NY, USA, **2006**, pp. 111–118.

[17] C. D. Geisler, *From Sound to Synapse: Physiology of the Mammalian Ear*, Oxford

University Press, **1998**.

[18] Y. Xie, T.-H. Tsai, A. Konneker, B.-I. Popa, D. J. Brady, S. A. Cummer, *Proc. Natl. Acad. Sci.* **2015**, *112*, 10595.

[19] D. El Badawy, I. Dokmanic, *IEEEACM Trans. Audio Speech Lang. Process.* **2018**, *26*, 2436.

[20] T. Jiang, Q. He, Z.-K. Peng, *Phys. Rev. Appl.* **2019**, *11*, 034013.

[21] D. El Badawy, I. Dokmanić, M. Vetterli, in *Latent Var. Anal. Signal Sep.* (Eds.: P. Tichavský, M. Babaie-Zadeh, O.J.J. Michel, N. Thirion-Moreau), Springer International Publishing, **2017**, pp. 89–98.

[22] S. Foucart, H. Rauhut, *A Mathematical Introduction to Compressive Sensing*, Springer New York, New York, NY, **2013**.

[23] R. Baraniuk, *IEEE Signal Process. Mag.* **2007**, *24*, 118.

[24] L. R. Rabiner, *Proc. IEEE* **1989**, *77*, 257.

[25] P. A. Deymier, *Acoustic Metamaterials and Phononic Crystals*, Springer Science & Business Media, **2013**.

[26] R. V. Craster, S. Guenneau, *Acoustic Metamaterials: Negative Refraction, Imaging, Lensing and Cloaking*, Springer Science & Business Media, **2012**.

[27] D. J. Brady, *Optical Imaging and Spectroscopy*, John Wiley & Sons, **2009**.

[28] Y. Chen, H. Liu, M. Reilly, H. Bae, M. Yu, *Nat. Commun.* **2014**, *5*, 5247.

[29] X. Zhu, B. Liang, W. Kan, Y. Peng, J. Cheng, *Phys. Rev. Appl.* **2016**, *5*, 054015.

[30] T. Jiang, Q. He, Z.-K. Peng, *Appl. Phys. Lett.* **2018**, *112*, 261902.

[31] R. A. Horn, C. R. Johnson, *Matrix Analysis*, Cambridge University Press, **2012**.

[32] A. V. Oppenheim, R. W. Schafer, *Discrete-Time Signal Processing*, Pearson, Harlow, **2014**.

[33] Y. C. Pati, R. Rezaiifar, P. S. Krishnaprasad, in *Proc. 27th Asilomar Conf. Signals Syst. Comput.*, **1993**, pp. 40–44 vol.1.

[34] J. A. Tropp, A. C. Gilbert, *IEEE Trans. Inf. Theory* **2007**, *53*, 4655.

[35] H. Hotelling, *J. Educ. Psychol.* **1933**, *24*, 417.

[36] J. C. Middlebrooks, D. M. Green, *Annu. Rev. Psychol.* **1991**, *42*, 135.

[37] P. Warden, *ArXiv180403209 Cs* **2018**.

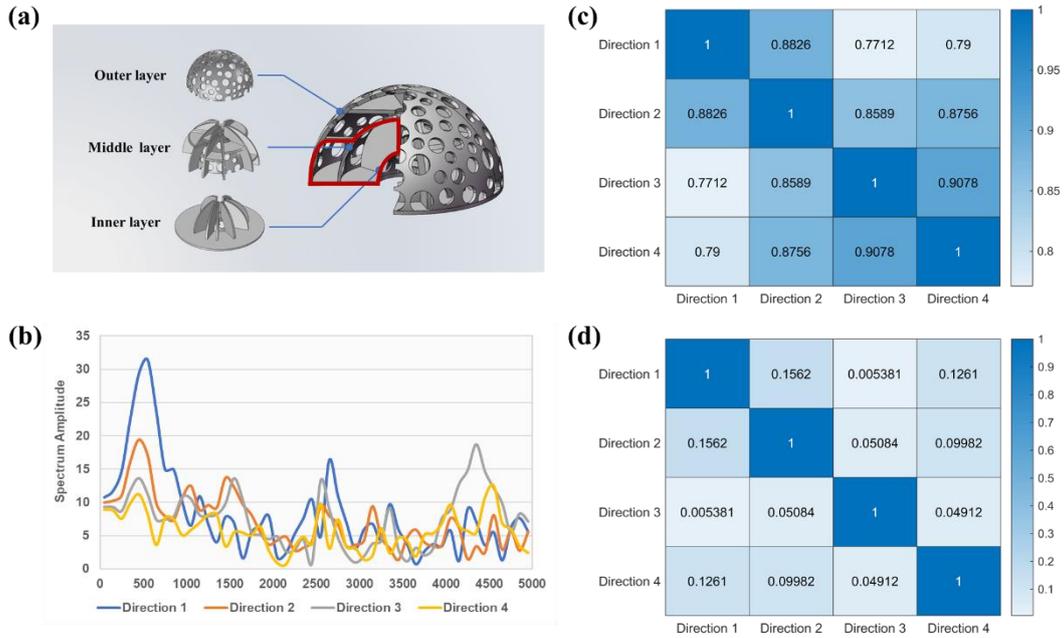

**Figure 1.** Model of the 3D ME. (a) Schematic view of the 3D ME: outer layer, middle layer, and inner layer. (b) Simulated frequency responses of the ME in four different directions. (c) The coherences between the 4 directions before VSPCA. (d) The coherences between the 4 directions after VSPCA.

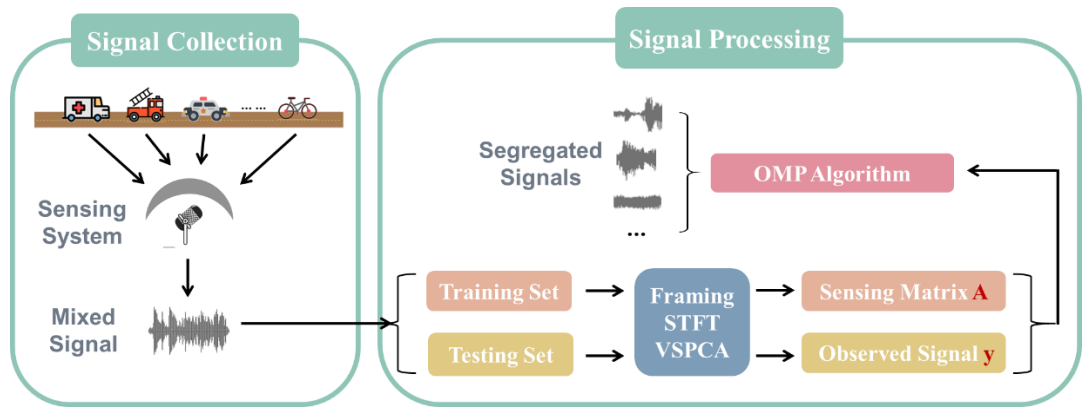

**Figure 2** Schematic of data collection and processing of the MSLS. The procedure of data collection is shown in the left frame and the procedure of signal processing is shown in the right frame.

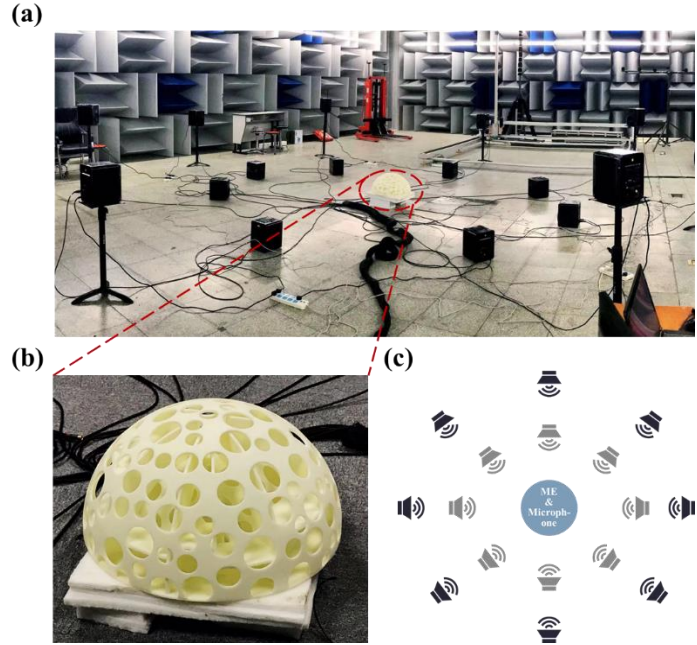

**Figure 3.** Measurement performed in a semi-anechoic room. (a) Photo of the experimental setup in the chamber. (b) Enlarged photo of the ME. A microphone is placed in the inner center of the ME. (c) Schematic of the setup: the ME and microphone are placed at the center, surrounded with 16 speakers.

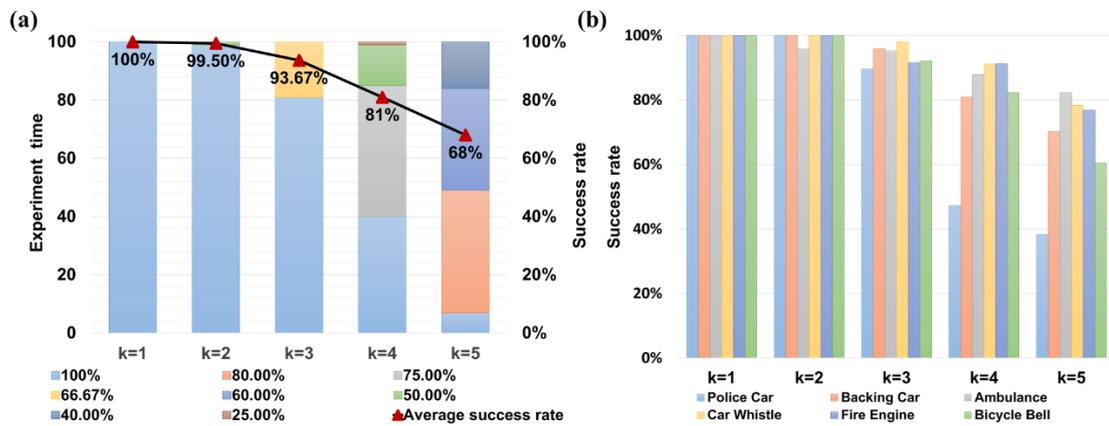

**Figure 4.** The results of the listening tests in the street scenario. (a) The results organized by the number of activated sources $k$. The success rate for each experiment is represented by different colours. For each $k$, the average success rate is calculated and represented by a red triangle. (b) Detailed results of the 6 kinds of signals in the street scenario. Each color represente a different audio content.

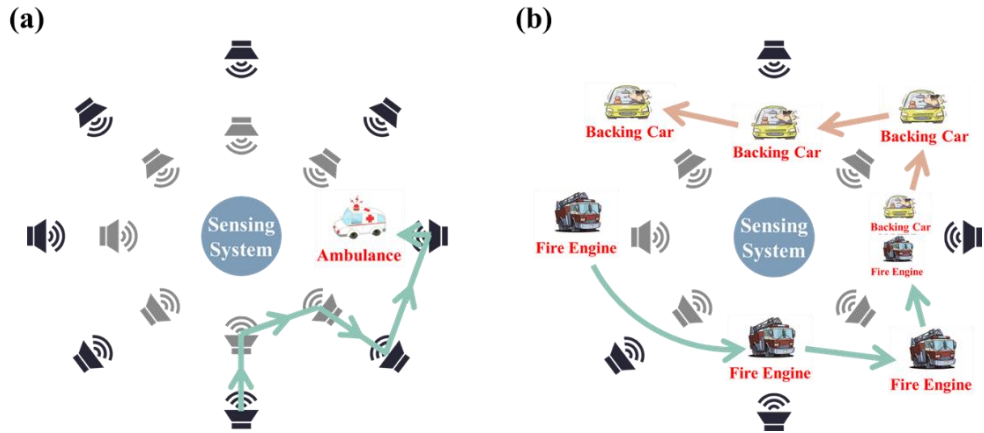

**Figure 5.** Listening tests of source identification and tracking. (a) The trajectories of a moving source of a ambulance. (b) The trajectories of two moving sources of the backing car and the fire engine. Detailed results are recorded in Supporting Videos 3 and 4.

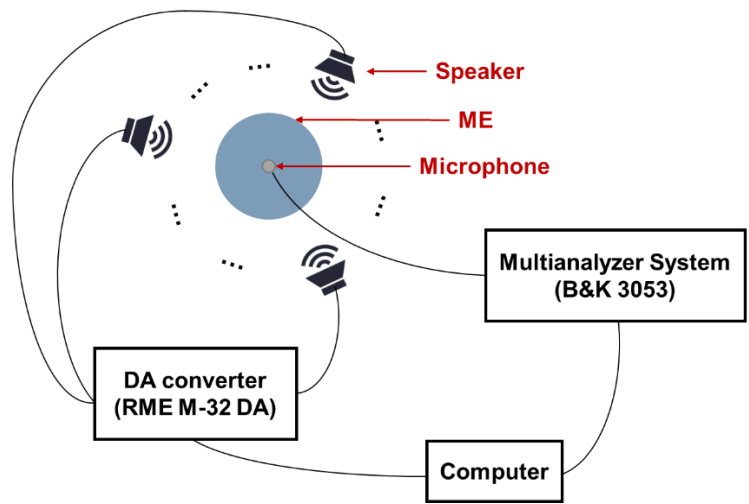

**Figure 6.** Schematic of the system layout.

|  | Home | Animal farm | Speech | Concert | Commands |
|---|---|---|---|---|---|
| **k=1** | 100.00% | 100.00% | 100.00% | 100.00% | 100.00% |
| **k=2** | 96.50% | 97.00% | 97.50% | 96.50% | 96.25% |
| **k=3** | 91.00% | 91.67% | 91.67% | 93.33% | 84.67% |
| **k=4** | 81.00% | 86.75% | 81.00% | 77.75% | 79.38% |
| **k=5** | 77.60% | 77.80% | 69.40% | 68.60% | 73.70% |

**Table 1.** The results of listening tests base on other datasets: home, animal farm, speech, concert, and commands.